\documentclass{article}
\usepackage{amsmath}
\usepackage{amsfonts}
\usepackage{amssymb}
\usepackage{latexsym}
\usepackage{verbatim}
\usepackage[title]{appendix}
\usepackage{epsfig}
\usepackage{comment}

 \usepackage{eso-pic}

\title{Quantifying Shannon's Work Function for Cryptanalytic Attacks}

\author{R. J. J. H. van Son\\
Netherlands Cancer Institute, Amsterdam and ACLC, University of Amsterdam\\
R.J.J.H.vanSon@gmail.com}

\begin{document}


\maketitle
\newpage

\begin{abstract}
\noindent
Attacks on cryptographic systems are limited by the available computational resources.
A  theoretical understanding of these resource limitations 
is needed to evaluate the security of cryptographic primitives and procedures. 
This study uses an Attacker versus Environment game formalism based on 
computability logic to quantify Shannon's work function and evaluate resource use in cryptanalysis. 
A simple cost function is defined which allows to quantify a wide range
of theoretical and real computational resources. With this approach the use of 
custom hardware, e.g., FPGA boards, in cryptanalysis can be analyzed. 
Applied to real cryptanalytic problems, it raises, for instance, the expectation
that the computer time needed to break some simple 90 bit strong cryptographic 
primitives might theoretically be less than two years.

\noindent
keywords: computation, cryptanalysis, computational complexity
\end{abstract}

\newpage
\section{Introduction}

There have been many examples where the ongoing increase in computer speed and 
capacities have made previously secure cryptographic systems vulnerable to brute 
force attacks. This perpetual weakening of cryptographic systems due to the 
progress in computer hardware has been incorporated in rules of application. 
For instance, NIST in the USA publishes elaborate rules about the phasing out 
of shorter (weaker) keys and algorithms over time \cite{Polketal2006,BarkerNIST2004}. 
However, those rules seem not to be based on a theoretical
understanding of the availability of computational resources, but more on a 
historical trend in technical progress (e.g., Moore's law \cite{tuomi2002lives}).

It is still difficult to reliably estimate the computational efforts needed to compromise a 
cryptographic system, i.e., Shannon's cryptanalysis work function
\cite{Shannon1949}. 
Many studies and applications go for ultimate security by aiming for $2^k$ operations, with 
$k\ge 128$, to put brute force attacks out of reach for the foreseeable future. Others use 
general purpose, off-the-shelf, computers as benchmarks. Both approaches 
have limitations. Long keys imply costly hardware and long computations and often do not describe
real life use, e.g., cost optimization for time-limited secrets. On the other hand, general 
purpose office and home computers are not necessarily 
very efficient for breaking codes and will almost certainly underestimate contemporary 
hardware capabilities \cite{Son0911-5262}.

The error to think that an off-the-shelf general purpose CPU for an office 
computer is an efficient device to recover cryptographic keys and passwords or
break cryptographic codes, is a common one. Expressions like
\textit{Calculating X took Y hours on a Z-level computer} are very 
often encountered. As a result, there seems to be general surprise every time
it is shown that low-cost, specialized processors can outperform general office
CPUs. For example, even though the idea might not have been new \cite{CookBK05}, there 
was again alarm in the media when in 2007 a Russian software company, Elcomsoft, filed for a 
US patent for a technique to use low cost standard graphics cards to recover passwords
\cite{HelpNetSecurity2007}.

Although the problem mentioned above is more generally seen in complexity and game theoretical 
analysis, it's practical importance is most acute for cryptanalysis and digital security.
Many security policies rely on cryptographic systems as a crucial element.
The difficulty with studying vulnerabilities in cryptography is their theoretical 
status. The most interesting vulnerabilities in cryptographic systems are generally 
untested and the cost of a theoretically possible attack is therefore very difficult to
estimate. Even though there is a good mathematical understanding of how
cryptographic systems can be compromised, there is no consensus about a
formalism in which the resources needed can be formally described and
quantified.

This study uses a general formalism for quantifying computational resources 
which was proposed in \cite{Son0911-5262}. This formalism defines resource use 
both on a symbolic level and on real hardware. The relevant parts 
of the model will be repeated here to make the current study self contained.
The model will then be tailored to quantifying the cryptanalysis 
work function of Shannon \cite{Shannon1949} which aligns very closely to problems 
in game theory, e.g., the computational Nash equilibrium 
\cite{Halpern-2008,HalpernPass-2008}, and algorithmic complexity theory with space 
and time bounded automata \cite{daylight-2008}.

Section  \ref{sectCryptGames} presents a summary of the model from \cite{Son0911-5262}
adapted to cryptanalysis. The use of the model will be illustrated 
on existing hardware products. In section \ref{SectBudget}, the model is applied to 
some examples from the cryptanalysis literature. The results are discussed in section
\ref{SectDiscussion}.

\section{Cryptanalytic attacks as games}\label{sectCryptGames}

Cryptanalytic attacks are interactive procedures where a cryptographic system
is attacked using computational resources to compromise protected
information. 
It is assumed that the attacker can only use algorithmic 
procedures and computers. Such an attack can
be emulated as a game by a collection of Turing Complete devices 
\cite{Shannon1949,NeumanMorgenstern1947}. For their mathematical
convenience, Universal Turing Machines (UTM) will be used to illustrate the
formalism \cite{Turing1936CNA}, but the results hold for all such devices.
Cryptanalytical attacks are problems of computability. 
This model of
cryptanalytic attacks fits the theoretic framework 
of \textit{computability logic} 
\cite{Japaridze03,japaridze2004,japaridze2005,GiorgiJaparidze08012006}.

In computability logic, computability is defined in terms of games.
The ``computer'', or Attacker, plays against the Environment and ``wins'' if it
can complete
the requested computation successfully. Computability logic tries to be
a complete logic of interactive computing. This study only refers to some
general aspects of computability logic. The reader can consult Japaridze
\cite{japaridze2005,GiorgiJaparidze08012006} and the references therein for
extensive descriptions of the theory.

In short, the Attacker can play a game against the Environment on one or more
``boards'', in parallel. This study will restrict itself to a \textit{Hard Play} model
of deterministic static
games \cite{japaridze2005}. That is, only purely algorithmic and reproducible
games are considered where the speed of the moves is not relevant.
The environment can execute
any number of moves for any single
computational step of the System. In practice, these two conditions, a
Hard Play model and static games, do not restrict the Attacker. They
just prescribe
that any attack strategy should involve a number of algorithmic steps and that
the Environment, which includes the complete universe, has unlimited
capacities for executing counter strategies. This model can be extended
to include probabilistic strategies.
In this framework, it is rather
straightforward to set up a model for a cryptanalytic attack
(c.f., \cite{NeumanMorgenstern1947,Shannon1949}).

\subsection{The Attacker model}

In the framework of computability logic, the Attacker is a collection of UTMs,
each with three tapes: a work tape, a
valuation tape, and a run tape. The
valuation tape is supposed to contain the game specific parameters supplied by
the Environment, whereas the work tape will be initialized with a program to
load and play games from the valuation tape. A more general interpretation of
the valuation tape is that it contains any public information outside the
control of the Attacker (for a more extensive description see \cite{Son0911-5262}).

The run tape contains the moves of the Attacker and the Environment. In the
current framework, both the Attacker and the Environment write their moves onto
the run tape. The alphabet used on the run tape is prescribed by
the Environment.
The Attacker can only move the reading head forward on the run tape and visit each
cell only once. The Environment is
free to read the run tape in any direction as often as it wants, but can only write to empty
cells. To allow the ``access once'' restriction, all moves are written as self delimited
or fixed length strings onto the run tape.

Scanning the run tape for moves of the Environment is a computational
cost that must be born by the Attacker. To minimize that cost, the moments
at which the Environment can write to a run tape are restricted.
The Environment will only write to a
run tape in response to a move of the Attacker. After the Attacker has written a
move to the run tape, it can enter a ``wait'' state and go to sleep.
Only then will the Environment write it's move or moves in a single self 
delimited or fixed length string to tape and wake up the Attacker who then 
can read the moves  and continue.
This interpretation of the
run-tape embodies the principle that the Attacker must actively query for
information from the Environment.

The Attacker can recruit as many UTMs as it wants by specifying
them on the run tape from any of the existing UTMs. The communication between
the Attacker UTMs is here
modeled by simply letting the work tapes overlap. Other solutions are possible.
Any newly instantiated UTM of the Attacker gets it's own run tape and a copy of
the valuation tape. 

Any request for a new UTM should consist of a full
description of the finite state machine, initial state, contents of the work
tape, position of the heads, and the overlap between  work tapes.
A new UTM is instantiated with the finite state machine
specified, the valuation and work tapes loaded and the work tape is stitched up
with the correct part of the requesting UTM's work tape. Then the UTM is put in
the initial state with the heads over the correct tape positions and started.

The possible moves of the Attacker can be divided into 4 classes: 
\begin{itemize}
\item General information requests
\item Structural requests
\item Encryption requests
\item Challenges
\end{itemize}
The meaning of the first is obvious. The second kind are requests to the Environment for 
new daughter UTMs or changes in the current UTM, e.g., releasing work tape memory.
In modeling a realistic attack, structural requests would (de-)commission computing resources.
Encryption requests implement the gathering of plaintexts and ciphertexts.

The cost of defeating a cryptographic system includes actually compromising it.
Challenges are Attacker initiated moves to prove it has won, i.e.,
succeeded in compromising the cryptographic system by actually executing and completing
the attack. A challenge could be to supply the current password, but could also 
demonstrate the ability to correctly pair cipher-
and plaintexts. Note that after every move of the Attacker, the Environment must make
a move, even if it is just a denial of the request.

This model of an Attacker is able to describe a large number of cryptanalytic
attacks. For instance, distributed attacks, both coordinated or not, known
ciphertext or plaintext attacks, and chosen plaintext attacks. Even attacks of
``security by obscurity'' systems could be studied by supplying a stochastic model
of information leakage to the valuation tape.

\subsection{Resource needs and cost of computation}\label{SectRequirements}

One problem with the above computational attack model is that most cryptographic
systems can be defeated by simple brute force attacks, e.g., just trying all
possible keys \cite{Shannon1949} or even simply trying all programs to crack
the encryption (c.f., algorithmic complexity \cite{LiVitanyi261084}). 
However, the security of cryptography
lies in the fact that performing computations has costs, and for a brute force
attack, these costs should be too high to be feasible \cite{Shannon1949}. But to
use these costs in a computational model, they should be made explicit.
In the remainder of this text, the model from \cite{Son0911-5262} will be
used to quantify computational costs. The relevant points will be described here.

A useful cost function for computations should follow some sanity conditions.
The definition should be applicable to both theoretical and real devices. The costs
should be cumulative and additive under appropriate conditions. The universal nature of 
computational devices should be mirrored in the existence of efficient emulation of one 
device on another one. Here an ``efficient emulator'' will be defined as any device that 
can emulate any computation on the target device with a cost that is a linear function 
of the original cost and number of steps in the computation (\cite{Son0911-5262}).

Starting with a purely theoretical device, a very simple cost function for a single 
UTM that agrees with all of the above conditions is
\begin{equation}\label{CostFunction}
C = \sum_{\lambda=1}^{\Lambda} I_{UTM}(\lambda)
\end{equation}
Where $C$ is the total cost of the computation which runs over $\Lambda$ steps.
$I_{UTM}(\lambda)$ is the information in bits stored in the UTM at step
$\lambda$. $I_{UTM}$ includes details about the internal structure 
of the device, e.g., action tables of a UTM. See \cite{Son0911-5262} 
for a discussion and proofs.

The definition of equation \ref{CostFunction} can easily be extended to other
computational devices (even neural wet-ware \cite{Son0911-5262}).
The only requirement is that the functionality of the device can be modeled
as a collection of interconnected and modular logical components, e.g., 
logical gates, finite state machines, or UTMs. The factor $I_{UTM}$ 
in equation \ref{CostFunction} will
be replaced by a factor $I_{Dev}$ which measures the number of bits
needed to identify the chosen device out of all the possible devices
(including all non-functional ones) that could have been constructed 
using the same basic components, 
plus the current state of these components.

For instance, the logical functions a modern CPU silicon chip can
perform are limited by the number of transistors it contains. The
size of $I_{CPU}$ would therefore be related to the number of ways
the transistors on it can be wired and how many states they can be in.
Note that in this description, no mention is made of the actual
physics of the components. That is, if the same range of logic 
functions could be performed using fluid valves or photonic 
switches, the same $I_{Dev}$ could result.

\begin{table}
\caption{Example processor characteristics.
Comp.: Parallel programmable components. \#Trans.: Indicative number of transistors.
Bytes/s: Resource size $I_{CPU}$ times cycles/sec from equation 
\ref{CostFunction}. Note that transistor counts are commercially sensitive
information which should be interpreted with extreme care. These numbers will also vary
widely between product versions.} 
\begin{tabular}{llrrrr}
Type &CPU & Comp.$^a$& Clock$^a$ & \#Trans.$^a$ & Bytes / s\\
\hline
GPU & ATI Radeon 5870 & 1712$^b$ & 850 MHz& 2.15$\cdot 10^9$ & 18.3$\cdot 10^{17}$\\
CPU & Intel Core Duo & 2 cores& 2.6 GHz & 291$\cdot 10^6$ & 7.57$\cdot 10^{17}$\\
FPGA & Xilinx Virtex-5 &  slices$^c$& &  &\\
 &\hfill{\footnotesize XC5VFX70T-2} & 11,200 &  249 MHz& 1.1$\cdot 10^9$ & 2.74$\cdot 10^{17}$\\
 &\hfill{\footnotesize XC5VLX30-3} & 4,800 &  251 MHz& 1.1$\cdot 10^9$ & 2.76$\cdot 10^{17}$\\
 &\hfill{\footnotesize XC5VFX70T-2} & 11,200&  277 MHz& 1.1$\cdot 10^9$ & 3.04$\cdot 10^{17}$\\
\end{tabular}\newline
{\footnotesize $^a$ Specifications as published in marketing materials and 
\cite{WikipediaCoreDuo,WikipediaTransistorCount,LoydCase2009}.
$^b$ The total number of stream processors, texture units, and render output units
\cite{LoydCase2009}.
$^c$ The Virtex-5 FPGA is organized in slices, with each slice containing four 6-input 
Look-Up-Tables (LUT) and four flip-flops \cite{NationalInstrumentsVirtex5,specification-virtex}}
\label{TabChipChars}
\end{table}%

\subsection{Relations with real hardware}\label{RealHardware}

The above theory on efficient emulators can be used to derive an estimate
of the computational capabilities of real hardware \cite{Son0911-5262}. 
As mentioned before, a CPU chip is characterized by a number of active
elements, transistors, and the connections between them. The whole CPU is 
run at a certain speed. The computational cost of running a certain 
computation on a CPU can therefore  be quantified as the number of steps
needed to complete the computation times the information frozen into the
chip design.

This exercise can also be done the other way around. 
First, the requirements for performing a basic computation in terms of electronic
circuits (i.e.,transistors) and number of steps are determined. Then, the
number of copies of the basic devices that fit on a silicon chip are determined.  After that,
the number and speed of the computations can be estimated, assuming 
state-of-the-art special purpose hardware could be used. It will not come as a surprise
to arrive at the conclusion that custom build electronics can often outperform
general purpose CPUs. 

Using the cost function of equation \ref{CostFunction}
and the device information content, $I_{Dev}$, can simplify this 
hardware analysis in many cases. It might obviate a
detailed analysis of the required circuitry and replace it with a less precise
but much more transparent calculation of comparable ``complexity''.

To make this analysis, a model is needed of the computational resources current hardware 
can deliver. A realistic model should take details of the limitations of chip design into 
account. In first approximation it is assumed that the maximal resources delivered by a CPU 
are proportional to the number of transistors. For the current study, a very crude model is assumed 
\cite{Son0911-5262}. For any given number of transistors on a chip, it is assumed that each 
transistor can be in one of two states (1 bit) and topological constraints limit the number of
different ways it can be connected to neighboring transistors to $\sim$100 (7 bit). In total, 
each transistor can thus be described with 1 byte.  Table \ref{TabChipChars} gives these numbers
for a few example processors.

This naive hardware model is illustrated below on some simplified cryptanalysis problems. 
The focus of the remainder of this section will be on compute-bound problems. The 
contribution of the memory components to the computations will be ignored in the analysis.

\subsubsection{Example: The EFF DES cracker}\label{SectEFFDESS}

In \cite{Son0911-5262}, the example of the EFF cracking the 56-bit single DES system in 
1998 \cite{EFF-DES,DEScracker:1998} is discussed. The challenge was to find the key that
could decrypt an unknown encrypted message.
From this example it is possible to get an
estimate of the number of transistors, and costs, needed to implement basic cryptographic
functions. The EFF succeeded in designing a search unit in silicon that could check 
a 56 bit DES key in 16 clock cycles \cite{EFF-DES,DEScracker:1998}. The EFF were able to fit 24 
such search units onto a single chip containing around 10,000 transistors and use 
the units in parallel. 

So a 56-bit DES encryption unit plus comparator needed $\sim 420$ transistors and 
runs in 16 clock cycles. With an estimated $I_{Dev} \sim 8\cdot \#$Transistors  bit, this 
comes down to around 6,700 bytes in equation \ref{CostFunction} for checking a 
single 56-bit DES encryption+compare
(i.e., $8\cdot 16 \cdot 10^4/24$ bits, c.f., \cite{Son0911-5262}). This translates to 
$\sim$120 byte per bit key length if it is assumed that encryption effort scales linearly 
with key length.

For a brute force key attack, the average number of keys that have to be tested scales 
with $2^{k-1}$ for key length $k$. For this specific DES attack, the computational costs, 
$C_{DES}(k)$, needed to find a key of length $k$ then scale as:
\begin{equation}\label{DEScosts}
 C_{DES}(k) =  120 \cdot k \cdot 2^{k-1}\;\text{(bytes)}
\end{equation} 
This cost will rise for Triple-DES. Probably in the order of tripling of the cost, e.g., 360 instead
of 120 byte per bit key length.

\subsubsection{GPU chips and super-computers}\label{SectGPU}

A modern Graphics Processing Unit (GPU) chip, like the ATI HD Radeon 5870, contains 
around 2.15 billion
transistors and runs at a clock speed of 850 MHz \cite{LoydCase2009}. 
Such a processor handles computations at a cost of $\sim 18 \cdot 10^{17}$ 
bytes per second (table \ref{TabChipChars}).
If such a processor could be constructed to run as an efficient parallel DES key search engine, 
i.e.,like the EFF custom chips, it would be able to find a 56 bit DES key in 133 seconds 
on average.

To illustrate the capabilities of GPUs, the analysis is extended to a hypothetical encryption
method with the same features as the single DES encryption standard, $DES^*$.
This $DES^*$ system is a model of simple cryptographic primitives and encryptions.
The fictional $DES^*$ differs from real DES in that it allows variable key lengths.
For every key length, an EFF DES cracker setup can be constructed for this fictional 
$DES^*$ that scales like equation \ref{DEScosts} and uses 120 byte per bit key length 
to check a single key.

On a customized processor of this size and speed, finding a 64 bit $DES^*$ key would 
require, on average, around 11 hours, and a 72 bit $DES^*$  key less than 5 months. A 
dedicated 65k ($2^{16}$) processor cluster would find an 84 bit $DES^*$  key in 
around 10 days and a 92 bit key in around 8 years. A 96 bit $DES^*$  key would take 
such a cluster around 120 years (on average; 240 years worst case). For 
finding a 96 bit $DES^*$  key in less than two years average, the technology would 
have to speed up by a factor of 60. At the historical rate of progress of $I_{Dev}$, 
around 2.6 dB/year ($\approx 1.82$/year \cite{Son0911-5262}), this would take another 
7 year to achieve (but see \cite{tuomi2002lives}).

For comparison, the fifth highest entry in the November 2009 TOP 500 list of supercomputers,
the Tianhe-1 supercomputer at the National SuperComputer 
Center in Tianjin/NUDT, China, contains 4096 Intel Xeon E5540 processors (2.5GHz, 
$7.3\cdot 10^8$ transistors) and 1024 E5450 processors (3GHz, $8.2\cdot 10^8$ transistors) 
connected to 5120 ATI Radeon HD 4870 GPUs (650MHz, $9.6\cdot 10^8$ transistors)
with a grand total of over 98TB of memory \cite{StromGPU2009,Tianhe1,Valich2009}.
Together the processors deliver $1.3\cdot 10^{22}$ bytes/sec (ignoring memory). If such a machine 
would have been build as a dedicated $DES^*$  key searcher, it would be able to find an 
84 bit $DES^*$  key in 87 days, on average. The Tianhe-1 was build for close to
88 million USD \cite{Valich2009}.

If the cost of encryption of Triple DES is indeed only $\sim$3 times that of single DES, the above 
numbers are not comforting. Triple DES with 2 independent 56 bit keys (keying option 2) has a listed
key strength much less than the expected 112 bits \cite{lucks1998attacking,OorschotWiener90known}. 
NIST designates this keying option to have only 80 bits of security \cite{Polketal2006} and retires it
in 2010. A message  encoded with the equivalent of an 80 bit  DES key could theoretically be decrypted within a 
few days with a special purpose 65k processor cluster as described above. However, the known 
attacks, e.g., \cite{OorschotWiener90known,lucks1998attacking}, are more complex than mere 
Triple DES encryption, with important time versus memory trade-off relations. 
Therefore, a separate analysis would be needed to calculate the costs of breaking double-key Triple DES.

\subsubsection{A better fit with FPGA}\label{SectFPGA}

The preceding sections assumed that an attacker could design and produce large numbers
of special purpose CPU chips with state of the art semi-conductor technology to compromise 
cryptographic systems. In many situations, such a threat model is unrealistic. In such cases, 
a better model would assume that the attacker would use existing customizable products.
A popular product in this class is a Field Programmable Gate Array (FPGA), an integrated circuit designed to be configured by the customer or designer after manufacturing \cite{WikipediaFPGA}.

Large differences in performance between general purpose processors and specially programmed 
(FPGA) chips have been demonstrated in the context of public key block ciphers by Gligoroski et al.
\cite{Gligoroski0808.0247}. 
They compared software implementations on a dual core Intel Core 2 Duo CPU
with implementations on Xilinx Virtex-5 FPGA chips  (table \ref{TabChipChars}).

On an Intel Core Duo dual processor, encrypting a 160 bit block with their MQQ\footnote{There are
successful attacks known against MQQ which preclude its use in encryption \cite{PersonalGligoroski}.
This does not affect the computational properties discussed here.} algorithm
takes 80,105 cycles and decrypting takes 6,212 cycles (tables 7 and 8 in \cite{Gligoroski0808.0247}). Assuming 
the CPU is running at 2.6GHz, this translates to a throughput of, respectively,  5.19Mb 
and 67.0Mb per second. Encryption of a basic data block (64 bit) with 1024-bit RSA requires
119,800 cycles, decrypting 2,952,752 cycles on the CPU. Throughputs for RSA are then,
respectively, 1.39Mb and 56.4Kb per second.

The same MQQ algorithm had a corresponding throughput for encryption of 44Gb per second
when implemented on four 276.7MHz Xilinx Virtex-5 FPGAs and 399Mb 
per second for decrypting when implemented on a single 249.4 MHz Xilinx Virtex-5
FPGA. An implementation of 1024-bit RSA on a 251MHz Virtex-5 FPGA had a throughput 
of 40Kb per second (unspecified for encryption or decryption). The computational
resources consumed when encrypting or decrypting a single bit are compared in table 
\ref{TabCompRes}.

For comparison, results for AES-128 on 16 byte blocks were collected. On an Intel Core Duo
E6700 CPU, the throughput was 1Gbps \cite{JemMatzan2006}. Two different implementations 
on Virtex-5 boards 
achieved 4.1Gbps throughput \cite{bulens2008implementation} (unspecified Virtex-5 types, assumed
to be the same as for the RSA, updating the results, 3.8Gbs, reported in \cite{Gligoroski0808.0247}). 

Efficient use of hardware is determined by the fit between algorithm and the logic implemented in the
chips. Encrypting with MQQ is amenable to parallelization and fits very well on the Virtex-5 
\cite{PersonalGligoroski}. From table \ref{TabCompRes} it can be seen that encryption with MQQ will use
$\sim$5300 times more resources (cycles$\cdot$transistors, i.e., bytes) when computed on a general 
purpose CPU than on a dedicated FPGA. 
An increase in hardware efficiency by a factor of $\sim$5300 would translate in an additional 12 bits
key length that could be decrypted for the same ``costs''. On the other hand, decryption shows
only a modest increase in efficiency by a factor of $\sim$16.

\begin{table}[b]
\caption{Computational resources consumed (bytes) when encrypting or decrypting 1 
bit using the MQQ based algorithm ($n$=160)\cite{Gligoroski0808.0247}, 1024-bit RSA
\cite{Gligoroski0808.0247}, and AES-128 \cite{JemMatzan2006,bulens2008implementation}. 
See table \ref{TabChipChars} for hardware specifications. The RSA results 
for the Virtex-5 combine encryption and decryption. See text for details.}
\begin{tabular}{lrrrrr}
                 & MQQ        &            &1024 RSA & &AES-128\\
                 & encryption & decryption &encryption & decryption & both\\
\hline
Core Duo & 146 GB& 11.3 GB & 272 GB$^a$ & 6.71 TB$^a$ & 379 MB\\
Virtex-5 & 27.5 MB$^b$  & 687 MB$^c$ & -$^d$ & 6.9 TB$^d$ & 67.3 MB\\
\end{tabular} \newline
{\footnotesize $^a$ per core \cite{PersonalGligoroski}. $^b$ four Virtex-5 XC5VFX70T-2 at 277 MHz.
$^c$ one Virtex-5 XC5VFX70T-2 at 249 MHz. $^d$ one Virtex-5 XC5VLX30-3 at 251 MHz, unspecified combined 
results for encryption and decryption were given.}
\label{TabCompRes}
\end{table}

Another algorithm, 1024-bit RSA, can hardly be parallelized and shows no real efficiency difference 
between CPU and FPGA. The AES-128 results are in between,
with a five time increase in efficiency between FPGA and general purpose CPU (assuming single core use).

The differences between the cases in table \ref{TabCompRes} raises the question of how the efficiency
gains can be understood. The large gains for the encryption using the MQQ algorithm implemented on
the Virtex-5 FPGA were derived from the ability to implement the steps of the algorithm in a pipeline that
could output one encrypted data block per clock cycle \cite{PersonalGligoroski}. Obviously, a 
tailored parallel pipeline approach is not possible with the fixed logic of a general purpose CPU. 
As illustrated by table \ref{TabCompRes}, such dramatic increases using FPGAs might be uncommon.

\section{Adversaries on a budget}\label{SectBudget}

A really Universal UTM can crack any cryptographic system that is based on
secret information that is less complex than the message. This can be done by
iterating over all programs and select the one that decrypts the message first.
In a secret key based system, it can be done by a brute force attack iterating
over all keys. However, brute force strategies can take more time and matter
than are available in the universe (c.f.,
\cite{SLloyd-2000_Limits,SLloyd02_Capacity,SLloyd05_quant-ph0501135}).
Therefore, a meaningful way is needed to limit the power of the Attacker without
losing the theoretical power of the UTM. The Attacker needs resources to perform
the required computations. Resources are understood in the sense of
\cite{GiorgiJaparidze08012006,Son0911-5262}. The resources are supplied by 
the environment on a request basis.

With a cost function to quantify computational needs in place, meaningful limits
can be placed on the Attacker. A budget is allocated to the Attacker, and before
every step in the computation, the resource costs of that computation step are
subtracted from the budget. If the budget becomes depleted, the Attacker loses.
The size of the smallest budget for which the Attacker can win the challenges
before the budget is depleted can be considered the strength of the
cryptographic system under study. It is obvious that a fully universal UTM is
regained in the limit of an infinite budget.

An intuitively meaningful way to set a budget is to calculate the computational
cost of testing all possible keys. So if testing one key costs $C_{key}$,
testing all keys of length $k$ bits will cost $C_{key} \cdot 2^{k}$, as
expected. To assist in book keeping, the Attacker can request the current size
of it's budget on the run tape. The valuation tape contains the information about
the resources available from the environment. For instance, in situations where
the Attacker does not have to design a computer system from scratch, the
valuation tape might contain a catalogue of available computer systems.

To illustrate the use of the above theory, a few cryptanalytical cases from the
literature are presented. Attention will be focussed on non-interactive
cryptanalysis. A full account should also address the interactive gathering of
information, e.g., differential cryptanalysis.

\subsection{Challenges: One-Time Pad example}\label{OTPchallenges}

Modeling cryptanalytical attacks as games enforces an explicit definition of
the conditions under which the Attacker wins. The computability logic model
described here defines winnability as the ability of the Attacker to succeed at
a number of predefined challenges. These challenges can be interactive.

For instance,
in most cryptographic systems, the ability to guess whether a known message
has been communicated would be a serious vulnerability. In the formalism
presented here, such knowledge could be formalized as being able to guess
above chance which ciphertext encodes a given plaintext.

As an example, suppose the challenge is to exploit a vulnerability in a One-Time
Pad (OTP) implementation where each plaintext is XORed (eXclusive OR) with a
unique sequence of random bits. The Attacker presents two self delimited
plaintexts on the run tape. The environment answers with a self delimited 
ciphertext that encrypts one of these plaintexts. The environment can pad the
shortest plaintext to the length of the
longest before encryption. The Attacker then tells which ciphertext was encrypted. 
If the Attacker can guess the correct plaintext above chance, the
Attacker wins. The threshold of proof can be put at any convenient level.

The attack strategy would then be to request encryptions of known or chosen
plaintexts. The One-Time pad bit strings are available for analysis after
removing (XOR-ing) the known plaintexts from the ciphertext. If some statistical
deviation from a pure, uncorrelated, uniform distribution can be detected in the bit strings,
the challenges can in principle be won. Simply chose the ciphertext that XORed
with the plaintext shows the anomaly.

As the OTP is proven secure \cite{Shannon1949}, the challenges are only winnable
if the (long) keys are not completely random, e.g., when using an insecure
Random Number Generator (RNG). An Attacker model might include a simulation
of compromising a
RNG as in, e.g., \cite{Gutterman1130388,kelsey98cryptanalytic}.
By varying the challenges between
ciphertext only, plaintext chosen by Environment, and plaintext chosen by Attacker
the effects of different security policies can be evaluated. For instance, the costs
and benefits of preventing guessing plaintexts can be compared to those of periodically
reseeding the key generator and redistributing new keys \cite{kelsey98cryptanalytic}.

Occasionally, the security of the OTP against cryptanalysis is questioned, as in 
\cite{Wang:cs0709.4420,Wang:cs0709.3334}. The formalism presented here can help to 
evaluate whether and how a vulnerability, if any, can be exploited.
For instance, from the analysis presented in \cite{Wang:cs0709.4420,Wang:cs0709.3334}
it is not clear how a chosen plaintext challenge as presented here can be won,
i.e., whether there is a vulnerability at all.

\subsection{Dictionary attacks and time versus memory trade-offs}

There exist methods to efficiently pre-calculate dictionaries with stored
ciphertext/key pairs to amortize the cost of encryptions over many different key
attacks \cite{an-Ping:cs0710.2970,Oechslin03}. To evaluate their threat, it is
necessary to estimate the resources needed to construct and operate such a
dictionary.
Constructing a table of Rainbow chains or a dictionary of
encryptions is equivalent to doing a brute force key search and requires the
same effort \cite{an-Ping:cs0710.2970,Oechslin03}.
The new question is how much resources are needed to use the dictionary
after it has been created.

For simplicity, assume a key size of $k$ and an ordered $(Ciphertext_i, Key_i)$
dictionary with $L = 2^{k-\epsilon}$ encryptions of a $3k$ long plaintext $X_0$
as in \cite{an-Ping:cs0710.2970}. The factor $\epsilon$ determines the fraction
of keys in the dictionary as $2^{-\epsilon}$. With these numbers, the size of
the dictionary is $D = 4kL$. According to \cite{an-Ping:cs0710.2970} it takes
at most $3k(k-\epsilon)$ comparisons to find an encryption in the dictionary,
but $k-\epsilon$ comparisons seems a more conservative choice. For $k=56$ and
$\epsilon = 6$, the size of the dictionary is $D = 4\cdot 56 \cdot 2^{50}
\approx 2.5\cdot 10^{17}$ bits,
or $3.1 \cdot 10^{16}$ bytes, and the expected number of comparisons per lookup
becomes $50$.

In the ideal case, every comparison is done in, say, two steps for a total of
$100$ steps per lookup. Assume that Attackers ``lease'' access to the
dictionary for each look-up, that is, there are no ``wait states'' and the resource
is in constant use by Attackers. The average
cost of a lookup is then $3.1 \cdot 10^{18}$ bytes, ignoring the small costs of the
comparisons themselves. The average cost of a discovered key would be around $2
\cdot 10^{20}$ bytes. Compared to the current scope of hardware, at $10^{18}$
byte/s for a single desktop system \cite{Son0911-5262}, this cost is unremarkable. 

The real point is not the
``computation'' or processing, but the required storage capacity of 31 petabyte
($31 \cdot 10^{15}$). This is around 15\% of the capacity of a large data-center
like Google's Googleplex facility, or a ``botnet'' of a few million computers
with some 10 GB each.  Such a resource would require parallel access through
many nodes, which would change the simple cost model above.
A botnet of this size would have to contain some 3 million
compromised computers with a real cost in the order of \$15 a piece, in 2007 
dollars, on the black market, or \$45 million in total \cite{Paxsonetal2007-1315292}. 
The combined value of the
encoded information must outweigh the costs of this set up to make this attack
worthwhile. The computational capacity of such a distributed data center or botnet,
with it's delayed response times, is obviously different from an integrated
desktop system.

This analysis shows that \textit{using} such a dictionary is, unsurprisingly, not
so much a computational as a storage problem. In this case, the maintenance of
such a large storage is much more a limitation than the duration of the computation.

\subsection{Pseudo Random Number Generator attacks: The TF-1 generator}

Pseudo-Random Number Generators (PRNGs) are important cryptographic
primitives that can be
vulnerable to their own types of attacks \cite{kelsey98cryptanalytic}.
PRNGs are used, for example, to generate the symmetric keys in public key
communication protocols like SSL (Secure Socket Layer protocol).
Their relative security, or lack thereof, is
strongly determined by the resources available to the Attacker
(e.g., \cite{kelsey98cryptanalytic}).

The Klimov-Shamir number generator TF-1 is analyzed by Tsaban
\cite{tsaban:0507063}. In short, for a word size $w$, this PRNG has an
internal state of size $4w$. The intended ``strength'' is $2^{2w}$
\cite{KlimovShamir04,tsaban:0507063}, i.e., $2w$ bit. However, Tsaban finds that the internal
state can be found in $16 \cdot 2^{1.5 w}$ elementary operations (i.e.,
$1.5w$ bit strength) after scanning  $2^w$ output words for a $0$ value
\cite{tsaban:0507063}. Each possible internal state can, on average, be checked
in 16 basic operations given a special $0$ value in the output.

The 16 operations needed to check the internal state are very basic. A DES
Cracker like search unit should be sufficient (see section \ref{SectEFFDESS}). 
The original DES Cracker search
unit used around 120 byte per bit key width. For the sake of argument, it is
assumed here that a comparable setup could be constructed that analyzes
the internal state again of the TF-1 number generator for 120 byte per bit in 
the reduced word size $1.5w$. 
Each basic operation should again need only a single clock cycle. For such a
system, the above analysis for the single DES cracker would still hold up to a
fixed factor (see sections \ref{SectEFFDESS} and \ref{SectGPU}).

An efficient setup with the complexity and speed of a ATI HD Radeon 5870 (see 
section \ref{SectGPU} and table \ref{TabChipChars}) would need under half a 
second to find the internal state
for a word width of $w = 32$ bit (48 bit strength) and less than five months
for a word width of $w = 48$ bit (72 bit strength), both on average (see table
\ref{TabTF-1}).
A cluster using 65 thousand such set-ups could finish a $w = 56$ bit word length
in ten days (84 bit strength). A theoretical $w = 60$ bit word length variant
(90 bit strength) could be expected to be broken in less than two years.
For word lengths of $w=64$ (96 bit strength), the time still runs into 120 years 
and remains elusive as Tsaban already notes \cite{tsaban:0507063}. 

\begin{table}[b]
\begin{center}
\caption{Expected times for finding the internal state of a TF-1 PRNG \cite{tsaban:0507063} 
using theoretically optimal custom CPUs with the complexity of an ATI HD Radeon 5870 
($1.83 \cdot 10^{18}$ Byte/s). See text for details.\newline
\textit{\#CPU}: number of CPU equivalents; \textit{\#values}: number of PRNG values needed to find special 0 value;
\textit{time}: expected time to find the internal state after finding the special 0 value.}
\begin{tabular}{lcrrl}
$w$ & strength (bit) & \textit{\#CPU} & \textit{\#values} & \textit{time} \\
\hline
32 & 48 & 1 & $2.1\cdot 10^9$ & 0.5 sec \\
48 & 72 & 1 & $1.4 \cdot 10^{14}$& 4.2 months \\
56 & 84 & 65,536 & $3.6 \cdot 10^{16}$ & 9.4 days \\
60 & 90 & 65,536 & $5.8 \cdot  10^{17}$ & 1.8 year \\
64 & 96 & 65,536 & $9.2 \cdot 10^{18}$ & 120 years \\
\end{tabular}
\label{TabTF-1}
\end{center}
\end{table}

The number of
output words needed to find a $0$ word can become unwieldy for the longer,
 $w = \{48, 56\}$, word lengths (see table \ref{TabTF-1}). For $w=48$, around 
 $2^{48-1}\approx 10^{14}$ output words have to be scanned for a 0 value. 
 That is around 40 hours at a billion ($10^9$) words per second (average). 
 For $w=56$ this would be a waiting time of 14 months. 
Note that originally, the intended strengths of
word lengths of $32$, $48$, and $56$ bit in TF-1 were, respectively, 
$64$, $96$, and $112$ bit.

An efficient attack of the TF-1 number generator would be to set up a cheap system to 
scan for 0-words storing a history of PRNG output and relevant data to compromise. 
Only after a 0-word has been encountered, the machinery to attack the cypher would be 
commissioned and the attack performed.  

No one has yet reported a DES Cracker like set-up for TF-1. So the above
calculations are based on the assumption that it could be possible to harness
the design complexity of a modern GPU for custom designed cryptanalysis
hardware.

The above analysis
allows to put a monetary number on the price to crack this specific PRNG. Users
of this algorithm can now judge themselves how much any adversaries would be
willing to pay for such a set-up and what the chances are of a version of the
algorithm that does \textit{not} need to find a $0$ word.

\section{Discussion and conclusions}\label{SectDiscussion}

Cryptanalysis promises to be a very fertile field for
developing insight into the quantification of computational resource needs.
A game theoretic view of cryptanalysis was introduced by Von Neumann and 
Morgenstern and later taken up by Shannon \cite{NeumanMorgenstern1947,Shannon1949}.
This study adopts this game approach and proposes to use \textit{computability logic}
\cite{japaridze2005,GiorgiJaparidze08012006,Son0911-5262} to rigorously define
Shannon's work function  \cite{Shannon1949}. In this approach, attack procedures are 
formulated in terms of computable functions \cite{Turing1936CNA}, the resources used, 
and also a full definition of the context of the attack.

Based on a few ``natural'' requirements, a simple formula for quantified
resources emerges as equation \ref{CostFunction} with the features of 
\textit{Memory} times \textit{Steps}, i.e., a dimension of bytes 
\cite{Son0911-5262}. This count includes the information
``frozen'' into the computational device itself, e.g., the UTM action table or the
components and connections of the CPU. By reducing silicon CPU complexity to
transistor connectivity and memory capacity, it is possible to roughly guess the
capacity of real hardware. 

Using the estimated hardware complexity of mass market processors as an
upper boundary, it is possible to estimate the limits of customized cryptanalytic
hardware. These limits can be used to understand historical cases, like the failure
of 56 bit DES encryptions \cite{DEScracker:1998}. 
These limits can also be used to predict the (theoretical) failure of modern
cryptographic primitives like the TF-1 PRNG with a theoretical strength
of $84$ and $90$ bit keys (intended strengths were originally $112$ and 
$120$ bits) \cite{KlimovShamir04,tsaban:0507063} as well as the efforts 
needed to actually effectuate the attacks.

It can be concluded that the general problem of quantifying computational 
resource use in interactive cryptanalysis attacks can be solved in a 
formalized setting. When used to formalize cryptanalysis, it becomes possible 
to quantify the cryptanalysis work function \cite{Shannon1949}. Even the 
computational costs of hypothetical attacks on cryptographic primitives can 
be estimated before they have to be demonstrated at great monetary cost. 

Examples show
that it would currently (2010) be feasible to build hardware that could break 
some 84 bit strength cryptographic primitives in mere days, and 90 bit 
strength primitives in less than two years.

\section{acknowledgment}
This project was made possible by grant 276-75-002 of the Netherlands
Organisation of Scientific Research (NWO)

\bibliographystyle{hacm}
\bibliography{CryptAttacks}

\end{document}